# Parametric generation of propagating spin-waves in ultra thin yttrium iron garnet waveguides

M. Mohseni, [1,*,**] M. Kewenig, [1,**] R. Verba, [2] Q. Wang, [1,3] M. Schneider, [1] B. Heinz [1,4] F. Kohl, [1] C. Dubs, [5] B. Lägel, [6] A. A. Serga, [1] B. Hillebrands [1], A. V. Chumak, [1,3] & P. Pirro [1]

[1]Fachbereich Physik and Landesforschungszentrum OPTIMAS, Technische Universität Kaiserslautern, Kaiserslautern, Germany

[2]Institute of Magnetism, Kyiv 03142, Ukraine

[3]Faculty of Physics, University of Vienna, Boltzmanngasse 5, A-1090 Vienna, Austria

[4]Graduate School Materials Science in Mainz, Staudingerweg 9, 55128 Mainz, Germany

[5]INNOVENT e.V., Technologieentwicklung, Prüssingstraße 27B, 07745 Jena, Germany

[4]Nano Structuring Center, Technische Universität Kaiserslautern, DE-67663 Kaiserslautern, Germany

We present the experimental demonstration of the parallel parametric generation of spin-waves in a microscaled yttrium iron garnet waveguide with nanoscale thickness. Using Brillouin light scattering microscopy, we observe the excitation of the first and second waveguide modes generated by a stripline microwave pumping source. Micromagnetic simulations reveal the wave vector of the parametrically generated spin-waves. Based on analytical calculations, which are in excellent agreement with our experiments and simulations, we prove that the spin-wave radiation losses are the determinative term of the parametric instability threshold in this miniaturized system. The used method enables the direct excitation and amplification of nanometer spin-waves dominated by exchange interactions. Our results pave the way for integrated magnonics based on insulating nano-magnets.

If the magnetization of a magnetically ordered system is driven out of its equilibrium state, the dynamics of the magnetic moments can form a spin wave (SW). [1,2] SWs, or their quanta, the magnons, carry a defined energy and momentum. Using SWs promises energy efficient devices to transport and process information on the basis of wave-based computing concepts in magnetic materials. [1-6] In this concept, information is encoded in the amplitude or the phase of the SWs. So far, several devices for this aim have been either proposed or realized, including magnonic crystals [7-8], directional couplers [5,9-14], majority gates [15], (de) multiplexers [16-17], transistors [18-19] and various logic elements. [20-24] Most of the mentioned devices utilize SWs in yttrium iron garnets (YIG) due to their ultra-low magnetic losses. [1, 25-28] However, downscaling those elements toward nanometer sizes is the basic requirement toward integrated magnonics where the magnons are controlled by magnons themselves. In principle, downscaling leads to a strong quantization of the energy levels, and consequently, to the appearance of distinct SW bands in the magnon spectrum. [28] This leads to suitable control of the SW wave vectors [29] since the excitation of SWs with a well-defined wave vector becomes easier, which is required for data processing using wave interference effects.

Parallel parametric pumping has proven to be a promising way to generate and amplify propagating SWs. [30-38] Parallel pumping can be realized when a pumping rf microwave (magnetic) field is applied parallel to the static magnetization of a uniformly magnetized film. Under this condition, the microwave photon with the frequency of $f_p$ splits into two counter propagating magnons with the frequency of $f_p/2$, under the conservation of energy and momentum. Parametric generation, which is the amplification of SWs from the thermal bath, is a threshold process. In general, the pumping threshold depends on the total magnon losses and the coupling of the pumping field to the magnonic system. [35] For localized parametric pumping, pumped magnons propagate away from the pumping source, and in addition to common relaxation losses, the SW radiative losses plays a substantial role. [38] Therefore, it is necessary to reassess the dominant mechanisms on the pumping threshold once the size of the pumping system is, for example, comparable to the wavelength of the SWs.

The field of nano-YIG magnonics is growing very rapidly nowadays [39-45], however, the knowledge of parametric generation of SWs in these systems is still lacking. Here, we present the experimental demonstration of the parametric generation of propagating SWs in a nanometer thick YIG waveguide (WG). Such a system exhibits a strong quantization in the lateral direction. Using optical detection via micro-focused Brillouin light scattering spectroscopy (μBLS), we demonstrate that the pumping field can excite distinct WG modes. Our experiments are analyzed using micromagnetic simulations and analytical calculations. We show how the SW radiation losses become determinative on the threshold of parametric instability in a YIG WG with a narrow pumping source.

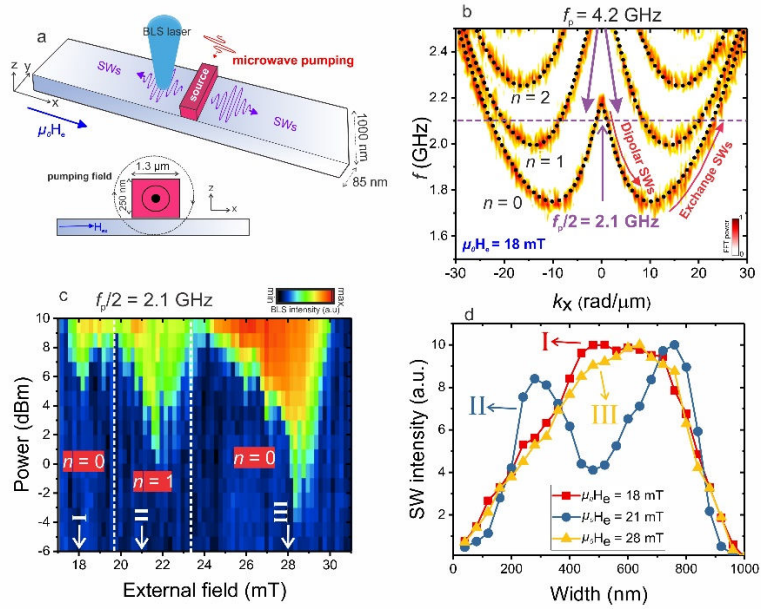

Figure 1: a) Schematic picture of the studied device and the microwave driven stripline on top of the waveguide; b) Magnon bandstructure of the system obtained from micromagnetic simulations (color plot) and analytically (dotted black lines) showing the waveguide modes which are labeled as $n$ = 0, 1, 2; c) Color coded spin-wave intensity measured at $f_p/2$ = 2.1 GHz as a function of the pumping power and the external field; d) spin-wave amplitude across the width of the waveguide when the external field is $\mu_0 H_e$ = 18 mT, $\mu_0 H_e$ = 21 mT and $\mu_0 H_e$ = 28 mT as labeled by I, II and III in Fig. 1c, respectively.

The investigated system here is a single nano-YIG WG with a thickness of 85 nm and a width of 1 μm. The nano-YIG films have been grown on a 500 μm thick Gadolinium Gallium Garnet (GGG) substrate with (111) orientation by liquid phase epitaxy [46]. The WGs were structured by using a metal hard mask technique with electron beam lithography and subsequent Ar+ ion beam etching. The stripline antenna with dimensions of 1.3 μm × 250 nm is placed on top of the WGs using electron beam lithography and a lift-off process. It consists of ~ 230 nm thick gold layer and 20 nm thick titanium layer for a better adhesion on the WG. The device is schematically shown in Figure 1a.

The SW dispersion relation of the longitudinally magnetized WG with an external bias field of $\mu_0 H_e$ = 18 mT is presented in Figure1b. The color plot shows the results from micromagnetic simulations

(thermal spectrum at 300 K) and the dotted lines are given by the analytical calculations [28]. The parameters, which have been considered for the analytical calculations and simulations, are as follows: $M_s$ = 140 kA/m, $A_{exch}$ = 3.5 pJ/m, $\alpha_G$ = 0.00025. For the simulations [47], the 40 µm long WG is divided into 2000×50×1 cells. This package uses the finite difference method (FDM) with a regular (rectangular) grid, which is significantly sipmler and faster than solvers based on the finite elements method (FEM). This is due to the fact that on a regular grid, the magnetostatic interactions can be efficiently calculated using Fast Fourier Transformation (FFT). Due to the lack of curved boundaries in our case, one cannot expect any additional computational errors, produced by the utilization of FDM in comparison to the FEM. The µBLS has been used for measuring the SW intensities of the pumped magnons [1].

In the presented configuration, the magnetization in the WG is parallel to the wave vector $k_x$. The quantization occurs along the lateral direction of the WGs and, therefore, WG modes appear in the spectrum. The WG modes are labeled by the number of the nodes in the dynamic magnetization in the width direction, and the spectrum of the first three modes, namely $n$ = 0, 1, 2 are shown in Figure 1b. Under this condition, the interplay between the dipolar and exchange contributions to the SW energy lead to the appearance of a local frequency minimum for each mode, the so-called band bottom. Thus, degenerate wave vectors of the same mode (or among different modes) exist for a certain range of frequencies. Indeed, increasing the wave vector for each mode leads to a smooth transition from the dipolar branch of the SWs spectrum (featuring large wavelengths) to the exchange branch of the SWs spectrum (featuring small wavelengths) as depicted in Figure 1b by red arrows.

We fix the microwave pumping frequency to $f_p$ = 4.2 GHz throughout the study. Figure 1c presents the color-coded SW intensity measured at the frequency of $f_p/2$ = 2.1 GHz as a function of the external field and the pumping power. The pumping threshold as a function of the bias field is known as the "butterfly curve" [35]. With respect to the external field, we observe three local minima of the threshold power for parametric SW generation. These lowest thresholds belong to field ranges around $\mu_0 H_e$ = 18 mT, $\mu_0 H_e$ = 21 mT and $\mu_0 H_e$ = 28 mT as labeled by I, II and III, respectively. The SW intensity profiles across the width of the WG can help to identify which SW eigen-modes are populated by the generated magnons. Thus, we set the field to $\mu_0 H_e$ = 18 mT, $\mu_0 H_e$ = 21 mT and $\mu_0 H_e$ = 28 mT and sweep the µBLS laser spot across the width of the WG. The results, which are shown in Figure 1d, indicate that the first SW mode, labeled as $n$ = 0 in Figure 1a is populated if the field is set to $\mu_0 H_e$ = 18 mT (red curve) and $\mu_0 H_e$ = 28 mT (orange curve). In contrast, the mode profile of the $\mu_0 H_e$ = 21 mT (blue curve) evidences that the second mode, labeled as $n$ = 1, is populated.

We now carry out numerical simulations in order to elaborate the wave vector of the parametrically generated magnons. Similar to the experiments in Figure 1d, we set the external field to $\mu_0 H_e$ = 18 mT, $\mu_0 H_e$ = 21 mT and $\mu_0 H_e$ = 28 mT and we drive the system with the frequency of $f_p$ = 4.2 GHz and an amplitude which is slightly above the threshold of the parametric pumping. The band structures of the magnons under pumping in the presence of the mentioned fields are presented in Figure 2. In particular, Figure 2a shows that magnons populate the dipolar branch close to the origin of the first mode if the field is set to $\mu_0 H_e$ = 18 mT. Increasing the field to $\mu_0 H_e$ = 21 mT leads to a transition of the generation from $n$ = 0 to the band bottom of the second mode, $n$ = 1, as shown in Figure 2b. A further increase in the external field to $\mu_0 H_e$ = 28 mT finally populates the band bottom of the first mode $n$ = 0 as presented in Figure 2c. Note that the band bottom of each mode exhibits a vanishing group velocity by definition.

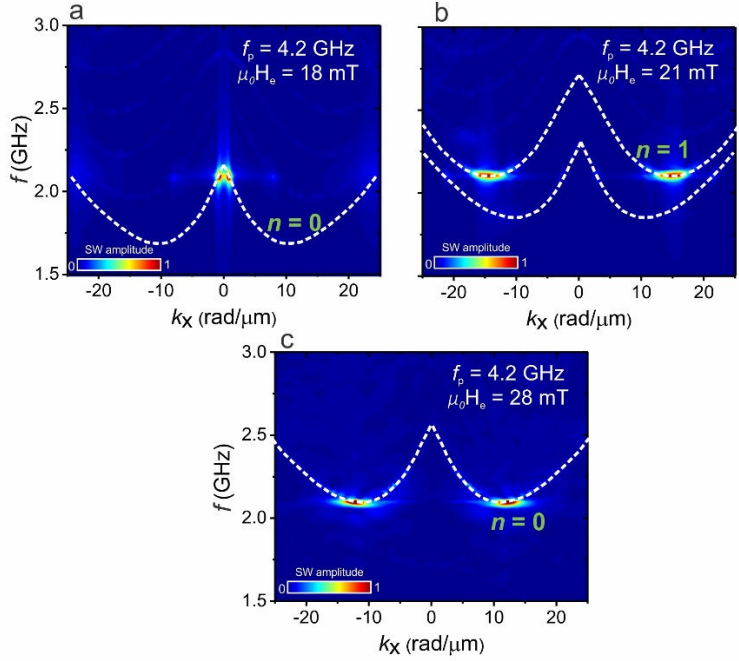

Figure 2: Color plot of the band structure of the system under pumping obtained from micromagnetic simulations; a) $\mu_0 H_e$ = 18 mT; b) $\mu_0 H_e$ = 21 mT; and c) $\mu_0 H_e$ = 28 mT. White dashed lines are showing the waveguide modes.

Such a mode transition of the pumped magnons from $n = 0$ to $n = 1$ via increasing the bias field is unusual since the first mode features the highest ellipticity and, consequently, highest coupling to the pumping field. Thus, one could expect that this mode possesses the lowest parametric instability threshold. [35] However, this argumentation is only valid as long as the radiative losses in the system can be neglected.

We carry out analytical calculations to find the parametric instability threshold of the respective SW modes. We only consider the involved modes in our observations which includes the first and second WG modes, $n = 0$ and $n = 1$, respectively.

We assume that the pumping field is uniform within the pumping area of the length Lp, and is zero outside of it. In the general case, the pumping threshold bth is determined from the following implicit equation [48,49],

$$\frac{\sqrt{(V_{kk}b_{th})^2 - (\Gamma_k - \alpha V_{kk}b_{th})^2}}{\Gamma_k - \alpha V_{kk}b_{th}} = -\tan[\frac{\sqrt{(V_{kk}b_{th})^2 - (\Gamma_k - \alpha V_{kk}b_{th})^2}L_p}{v_{gr}}] \qquad (1)$$

Here $\Gamma_k = \alpha_G \varepsilon_k \omega_k + \Delta\omega_{nu}$ is the SW damping rate, consisting of Gilbert losses ($\varepsilon_k$ is the coefficient, related to the averaged ellipticity of magnetization precession in the SW mode [50]) and nonuniform line broadening $\Delta\omega_{nu}$ (assumed to be negligible in our case). We note that the SW Gilbert damping rate $\alpha_G$ should not be confused with the nonadiabaticity parameter $\alpha$ (see below). Furthermore, $v_{gr}$ is the SW group velocity, and $V_{kk}$ is the modulus of the efficiency of the parametric interaction, also known as the coupling parameter, which in our geometry of parallel pumping is given by [31,35],

$$V_{kk} = \gamma \frac{|\langle m_y^2 - m_z^2 \rangle|}{4\langle m_y m_z \rangle} \qquad (2)$$

where $m_{y,z}(y)$ are the distributions of the dynamic magnetization (SW profile) and the symbols $\langle ... \rangle$ indicate averaging over the waveguide width. Finally,

$$\alpha = |b_{2k}/b_0| \qquad (3)$$

is the parameter of nonadiabaticity [49], determined via the spatial Fourier-harmonics $b_k$ of the pumping profile $b_p(x)$. In our case, it is equal to,

$$\alpha = |\text{sinc}[kL_p]| \qquad (4)$$

The pumping threshold is determined by the total losses in the system, which consist of relaxation losses and radiation losses (but, is not a simple sum of them), and by the efficiency of the parametric interaction. In the limiting case of a large pumping area with $L_p \gg v_{gr}/\Gamma_k$, radiation losses are negligible (and the nonadiabaticity also vanishes), and the threshold is equal to

$$b_{th} = \Gamma_k/V_{kk} \qquad (5)$$

In the opposite case, for a small pumping area when relaxation losses are negligible, the threshold is equal to [49]

$$b_{th} = \frac{v_{gr}}{L_p V_{kk}} \frac{\arccos \alpha}{\sqrt{1-\alpha^2}} \qquad (6)$$

In our case the parametric resonance condition can be satisfied simultaneously for two SW modes, namely $n = 0$ and $n = 1$, in the field range where both modes have solutions at $f_p/2$. Otherwise, this frequency degeneracy still exists between the dipolar and exchange branch of a single mode. Nevertheless, the parametric pumping excites the mode with the lowest pumping threshold. Excitation of another mode having larger threshold could happen at a pumping strength, which is significantly higher than this formal threshold, because the already excited mode increases the effective losses for any other mode [51-52].

The parametric interaction efficiency and relaxation losses are both dependent on the precession ellipticity $\mathcal{E} = m_{max}/m_{min}$. A larger precession ellipticity leads to an increase of both $\Gamma_k$ and $V_{kk}$. However, the ratio $\Gamma_k/V_{kk}$ is not constant. In principle, the ellipticity-related prefactor $\varepsilon_k$ of the damping rate increases from 1 for circular precession to higher values with increasing $\mathcal{E}$. On the other hand, $V_{kk} = 0$ for circular precession. If the precession ellipticity is constant across the WG width (i.e. if dynamic magnetization profiles are described by the same spatial function, $m_y = m_{y,0} f(y)$, $m_z = m_{z,0} f(y)$), then the ratio of relaxation losses to the parametric interaction efficiency is proportional to

$$\frac{\Gamma_k}{V_{kk}} \sim \frac{\mathcal{E}^2 + 1}{\mathcal{E}^2 - 1} \qquad (7)$$

This ratio monotonically decreases with the increase of the ellipticity (ratio changes from infinity for $\varepsilon = 1$ to 1 for $\varepsilon \to \infty$). Therefore, in the limit of negligible radiation losses, always the SW mode with the largest ellipticity is excited. In our geometry, the SWs in the dipolar branch always have larger ellipticity, because the dipolar interaction is anisotropic, while the exchange interaction is isotropic. In addition, a lower WG mode has a larger ellipticity in comparison to the higher WG modes. This is due to the presence of the higher nonuniformity of the SW profile across the WG width, which increases the dynamic dipolar fields in this direction. Thus, as presented in Fig. 3a, the threshold in the limit of a large pumping area is always smaller for the dipolar branch of the lowest SW mode which satisfies the resonance condition $f_k = f_p/2$.

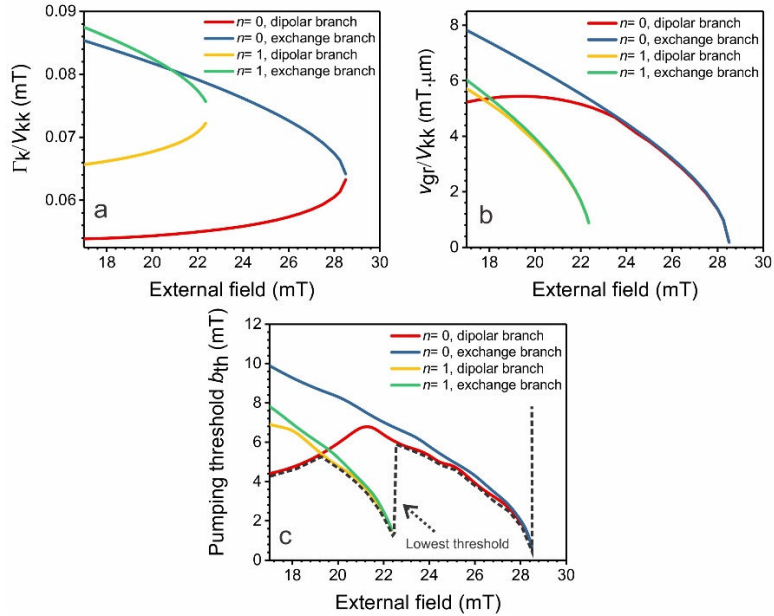

Figure 3: Pumping threshold field with respect to the external field obtained from analytical calculations; a) pumping threshold by considering only the Gilbert losses from equ.5; b) the ratio of $v_{gr}/V_{kk}$ which determines the radiation losses; and c) Pumping threshold by considering both Gilbert and radiation losses. Note the difference of the scales and values on the $y$-axis of all figures.

In the limit of small pumping lengths, the threshold is determined by the ratio of the SW group velocity to the parametric interaction efficiency. It is clear that this ratio is vanishingly small at the band bottom, where SW group velocity approaches to zero. That is why we observe an excitation of the $n = 1$ mode with low threshold when its bottom satisfies the resonance condition as shown in Figure 2b. In addition, as it is clear from Figure 3b, the ratio $v_{gr}/V_{kk}$ is smaller for dipolar branch of both modes. However, depending on the bias field, this ratio is smaller for the dipolar branches of different modes (yellow and red curves in Figure 3b). Noting, finally, that the nonadiabaticity parameter is larger for longer (dipolar) SWs and, thus, promotes excitation of these SWs, we conclude that in our geometry always dipolar SWs are excited.

Taking both losses and the nonadiabaticity into account leads to a threshold curve which is presented in Figure 3c. In our case, in which the pumping source length is $w = 1300$ nm, the radiation losses are determinative for the excitation threshold. This is clear from the comparison of the threshold value

(Figure 3c) with its relaxation contribution (Figure 3a). The theoretical curve in Figure 3c is in a very good agreement with our experimental observation of the butterfly curve as presented in Figure 1c. By comparing the experiments with the theory, one can point out that there is no abrupt increase of the threshold at the right side of the minima in the experimental curve, in contrast to the theoretical one. This is a consequence of the nonresonant excitation of the SWs near these minima [53-51]. Indeed, while the nonresonant excitation significantly increases the threshold, in a certain small range, this nonresonant excitation of a mode near the minima is more favorable than the resonant excitation of another mode.

We note that the ratio $v_{gr}/V_{kk}$, which is determining the radiation losses, is not always smaller for dipolar SWs in comparison to the exchange SWs. Depending on the WG geometry, the opposite situation can take place. Under this condition, the theory predicts that one can directly pump to the exchange branch of a given mode via changing the size of the pumping source. This means the transition between the excitation of the dipolar to the exchange SWs happens at a critical size of the pumping source. Under this condition, the contribution of radiation losses becomes sufficiently large to overcome *i*) the gain in relaxation losses and, *ii*) the higher parametric interaction efficiency of the dipolar SWs. Using this method, one could directly excite nanometer SWs in similar systems by designing the pumping source. On the other hand, using sufficiently large pumping sources leads to the absence of radiation losses and parametric excitation of only dipolar SWs. This will be similar to macroscopic YIG films with a hybridized spectra where the parametric amplifications using microwave resonators are expected to only excite the modes with the highest ellipticity, which is dominated by the dipolar interactions. [55]

We finally note that, for certain directions of the waveguide with respect to the YIG crystalline axes, the cubic anisotropy gives a non-vanishing contribution both to the SW frequency and to the parametric interaction efficiency, which is most pronounced at low external fields and small wave vectors. In our case, this contribution can be neglected safely, as follows from the comparison of the experimental results and calculated SW spectra.

In conclusion, we have investigated the parametric generation of propagating SWs in a nanometer-thick YIG waveguide. Using Brillouin light scattering microscopy, the parametric excitation of the two WG modes by a narrow pumping source has been demonstrated. Micromagnetic simulations clarified the wave vector of the generated SWs. Analytical calculations demonstrated which mode features the lower threshold and indicate that the SW radiation losses are the dominant term of the parametric instability threshold in this system. Using this knowledge enables the direct excitation and amplification of nanometer SWs in nano-YIG based devices. Furthermore, it introduces a way to amplify a selected SW mode in a carefully designed system. Our results provide an insight into nonlinear SW dynamics in insulating nano-magnets, aiming at further developments for designing new magnonic integrated architectures.

This project is funded by the Funded by the Deutsche Forschungsgemeinschaft (DFG, German Research Foundation) - TRR 173 – 268565370 (Project B01), and through the project DU 1427/2-1, by the Nachwuchsring of the TU Kaiserslautern, by the European Research Council Starting Grant No. 678309 MagnonCircuits, by the Austrian Science Fund (FWF) through the project I 4696-N, and by MES of Ukraine (project 0118U004007) are gratefully acknowledged. B.H. acknowledges support by the Graduate School Material Science in Mainz (MAINZ).

---

*mohseni@rhrk.uni-kl.de,

**Authors with equal contribution